\documentclass[]{mynature}
\usepackage{graphicx}
\usepackage{multirow}
\usepackage{multicol}
\usepackage{amsmath}
\usepackage{bibunits}
\usepackage{bm}
\usepackage{color}
\usepackage{soul}

 \defaultbibliographystyle{naturemag}

\newcommand{\beginsupplement}{%
        \setcounter{table}{0}
        \renewcommand{\thetable}{S.\arabic{table}}%
        \setcounter{figure}{0}
        \renewcommand{\thefigure}{S.\arabic{figure}}%  
     }

\title{Evidence for Coexistence of Bulk Superconductivity and Itinerant Antiferromagnetism in the Heavy Fermion System CeCo(In$_{1-x}$Cd$_x$)$_5$} % Max 15 words

\author{Ludovic~Howald$^{1,2,\star}$, Evelyn~Stilp$^{1,3}$, Pierre~Dalmas~de~R\'eotier$^4$, Alain~Yaouanc$^4$, St\'ephane~Raymond$^4$, Cinthia~Piamonteze$^2$, G\'erard~Lapertot$^4$, Christopher~Baines$^3$ \& Hugo~Keller$^1$}

\begin{document}

\newcommand{\CeCoIn}{CeCoIn$_5$}
\newcommand{\CeCoInCd}{CeCo(In$_{1-x}$Cd$_x$)$_5$}
\newcommand{\Tc}{$T_\textrm{c}$}
\newcommand{\TN}{$T_\textrm{N}$}
\newcommand{\muSR}{$\mu$SR}
\newcommand{\modif}[1]{\textcolor{red}{#1}}
\newcommand{\corr}[1]{\textcolor{red}{\st{#1}}}

\maketitle

\begin{affiliations}
 \item Physik-Institut der Universit\"at Z\"urich, Winterthurerstrasse 190, CH-8057 Z\"urich, Switzerland
 \item Swiss Light Source, Paul Scherrer Institut, CH-5232 Villigen PSI, Switzerland
 \item Laboratory for Muon Spin Spectroscopy, Paul Scherrer Institut, CH-5232 Villigen PSI, Switzerland
 %\item Institut Nanosciences et Cryog\'enie, SPSMS, CEA and University Joseph Fourier, F-38054 Grenoble, France
\item Universit\'e Grenoble Alpes, INAC-SPSMS, F-38000 Grenoble, France and CEA, INAC-SPSMS, F-38000 Grenoble, France
\item[$\star$] ludovic.howald@psi.ch
\end{affiliations}

\begin{abstract} %150 words  general introduction to the topic and as a brief, non-technical summary of the main results and their implications

%Unconventional superconductivity is found in systems close to phase instabilities. 
In the generic phase diagram of heavy fermion systems, tuning an external parameter such as hydrostatic or chemical pressure modifies the superconducting transition temperature. The superconducting phase forms a dome in the temperature--tuning parameter phase diagram, which is associated with a maximum of the superconducting pairing interaction.
Proximity to antiferromagnetism suggests a relation between the disappearance of antiferromagnetic order and superconductivity.
We combine muon spin rotation, neutron scattering, and x-ray absorption spectroscopy techniques to gain access to the magnetic and electronic structure of \CeCoInCd{} at different time scales.
Different magnetic structures are obtained that indicate a magnetic order of itinerant character, coexisting with bulk superconductivity.
The suppression of the antiferromagnetic order appears to be driven by a modification of the bandwidth/carrier concentration, implying that the electronic structure and consequently the interplay of superconductivity and magnetism is strongly affected by hydrostatic and chemical pressure.

\end{abstract}

\begin{bibunit} 
% 5000 words without methods

%1010 intro + 177 samples + 1425 magnetism/SC + 1274 XAS + 1004 discussion = 4890

Understanding the interactions between superconductivity and magnetism in heavy fermion systems is one of the greatest challenges of condensed matter physics, often presented as a key to unveil the mechanism of unconventional superconductivity. In the generic phase diagram of these systems (see typically Fig.~1), superconductivity arises in proximity to a second order magnetic phase transition, possibly leading to a quantum critical point (QCP)\cite{McMillan1968}. The identification and the understanding of QCPs play a major role in the search of the pairing mechanism. A softening of an excitation mode (phonon, magnon, ...) occurs at a QCP. Shall this mode be related to the pairing interaction, a maximum in the superconducting (SC) coupling would be observed\cite{Monthoux2007}. In any case, it appears rather probable that the same mechanism causes the suppression of the magnetic state and the rise of superconductivity.

The tetragonal system \CeCoIn{} has the highest SC transition temperature (\Tc$\simeq$2.3\,K) among the cerium based heavy fermion family\cite{petrovic2001}, and a pressure induced QCP at $p_c\simeq 0.4$\,GPa (red star in Fig.~1) as determined by SC parameters\cite{Howald2011b} (maximum SC coupling and pair breaking). Extrapolation of $p_c$ from the normal state is more difficult and has led to a value in the range: -0.5\,GPa to 1.3\,GPa\cite{Nicklas2001,Ronning2006,Zaum2011}. The antiferromagnetism, often believed to be at the origin of the QCP, can be revealed by different doping: (i) Cd or Zn at the In site\cite{Pham2006,Yokoyama2014}, (ii) Rh at the Co site\cite{Goh2008} or (iii) Nd at the Ce site\cite{Raymond2014}. Antiferromagnetism is also found by applying a large magnetic field in the $(a,b)$-plane\cite{Kenzelmann2008,Kenzelmann2010} ($\sim 10 - 11.5$\,T). 
The antiferromagnetic (AFM) structure was fully resolved in the case of CeRhIn$_5$ as incommensurate with in-plane moments mainly localized  on the cerium site forming a spiral structure in the $\vec{c}$-axis crystallographic direction\cite{Bao2000,Schenck2002}. In \CeCoIn{} neutron diffraction experiments report a commensurate order when the AFM phase is reached upon Cd or Rh doping\cite{Nicklas2007,OhiraKawamura2007} and an incommensurate one in an applied magnetic field\cite{Kenzelmann2008} or upon Nd doping\cite{Raymond2014}. 
The cadmium doped samples have the same phase diagram compared to the pure compound, beside a shift in the pressure scale. 
 They are therefore presented as ``equivalent'' to negative hydrostatic pressure\cite{Pham2006}. The ``equivalence'' can be understood in terms of a larger ionic or covalent radius for cadmium compared to indium. Cadmium atoms also contain an electron less than indium atoms and are therefore expected to increase the carrier concentration in the hole character bands of the Fermi surface of \CeCoIn . Similarly, hydrostatic pressure strongly modifies the electronic structure in \CeCoIn{} as shown by the increase of the number density in the superconducting phase \cite{Howald2013} and the decrease of the electronic specific-heat coefficient $\gamma$ \cite{Sparn2002} with pressure. These variations can be understood as a transfer of charge carriers between the different electronic bands under hydrostatic pressure. The two band hole sheets are the ones of strong 4$f$ angular momentum character believed to play an important role in \CeCoIn , both for superconductivity and antiferromagnetism\cite{Maehira2003}.

The suppression of the AFM phase potentially leading to the QCP and superconductivity in \CeCoIn{} is usually understood in term of ``delocalization'' of the Ce\,4$f$ electrons. In heavy fermions systems, at least three different mechanisms can ``delocalize'' the Ce\,4$f$ electrons: (i) the Kondo effect\cite{Kondo1964}, (ii) an increase in cerium valence\cite{Watanabe2011}, and (iii) the formation/broadening of an electronic band of partial Ce\,4$f$ character (Mott delocalization)\cite{deMedic2005}. Each of these effects can independently introduce the Ce\,4$f$ electrons into the Fermi surface. Note that Ce\,4$f$ electrons belonging to the Fermi surface (referred to as itinerant) via one of the previous effects can still be ``delocalized'' by the two other mechanisms. 
A variation of the Ruderman-Kittel-Kasuya-Yoshida (RKKY) coupling is unlikely to be the cause of the suppression of the AFM phase as no ferromagnetism was reported under higher pressures\cite{Knebel2004}.

An example of a local--itinerant transition takes place between the parent AFM system CeRhIn$_5$ in which most of the Ce\,4$f$ electrons do not participate to the Fermi surface and \CeCoIn{} in which most of the Ce\,4$f$ electrons participate to the Fermi surface. Such transition can be observed experimentally via a modification of de-Haas-van-Alphen frequencies between \CeCoIn \cite{Hall2001} and CeRhIn$_5$\cite{Alver2001}.
A comparison to the equivalent lanthanum system,  which has no 4$f$ electrons, leads to the same conclusion:
The volume of the Fermi surface is modified upon lanthanum doping in Ce$_{1-z}$La$_z$CoIn$_5$ and stays constant in Ce$_{1-z}$La$_z$RhIn$_5$\cite{Harrison2004}. 
The AFM transition in this family is however not directly related to the local--itinerant transition.
In the intermediate system CeRh$_{1-y}$Co$_y$In$_5$ a reconstruction of the Fermi surface occurs at $y=0.4$, associated with partial inclusion of the Ce\,4$f$ electrons into the Fermi surface for $y>0.4$, while antiferromagnetism persists up to $y\simeq 0.8$\cite{Goh2008}. Similarly, in \CeCoInCd{} the variation of the Fermi volume with $x$ is weak and comparable to the case of LaCo(In$_{1-x}$Cd$_x$)$_5$ at least up to $x=0.075$ which is already an AFM system\cite{Capan2010}. The suppression of antiferromagnetism must therefore be ascribed to a ``delocalization'' of itinerant Ce\,4$f$ electrons.

The isostructural $\alpha$--$\gamma$ phase transition of pure cerium shows strong analogies to the AFM--paramagnetic transition of heavy fermion systems\cite{Flouquet2005}. The large volume change of the unit cell at the $\alpha$--$\gamma$ phase transition is also attributed to one of the ``delocalization'' mechanisms discussed above\cite{Nikolaev2012}. The various proposed models are: (i) valence variation of $\approx 0.7$ (promotional model)\cite{Kornstaedt1982}, (ii) Kondo volume collapse model\cite{Allen1982} or (iii) formation of a band of partial Ce\,4$f$ angular momentum character  (Mott transition)\cite{Johansson1974}.

The question addressed through the experiments of this study is the origin of the suppression of the magnetic order in \CeCoInCd , that is, identifying the ``delocalization''  mechanism and its relation to superconductivity. In \CeCoInCd , the AFM phase was probed by muon spin rotation (\muSR ) and neutron diffraction. The \muSR{} technique also established the coexistence of antiferromagnetism with superconductivity, while the valence of cerium was determined by x-ray absorption spectroscopy (XAS). A comparison to metallic cerium allows resolving the nature of the transition. The combination of these three different experimental probes allows a deep understanding of the ground state of this system that cannot be achieved with a single technique.

\section*{Results}
\subsection{Crystals with negative hydrostatic pressure.}

\begin{figure}
\begin{center}
\includegraphics[width=85mm]{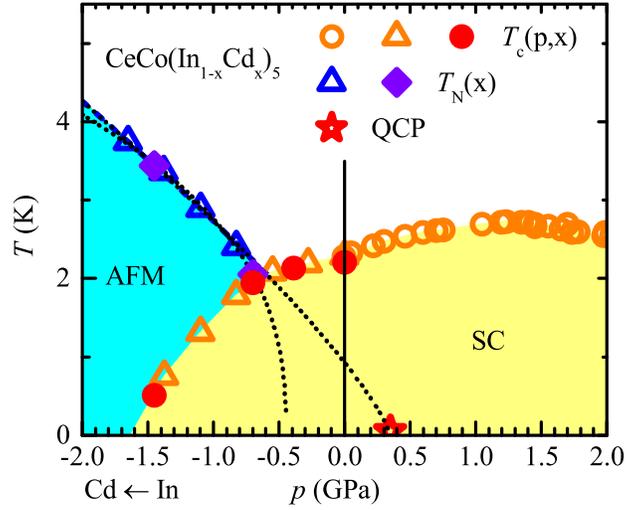}%width=0.5\textwidth
\end{center}
\caption{\label{Cp_phasediagram} \textbf{Phase diagram of \CeCoInCd} with the equivalent ``negative'' chemical pressures adapted from Ref.~\citen{Pham2006} (empty blues and oranges triangles) and positive hydrostatic pressures from Ref.~\citen{Knebel2004} (empty oranges circles). The red star represents the position of the putative QCP\cite{Howald2011b}. The black dotted lines indicate extrapolations of the AFM state as expected in absence of superconductivity. In one case the QCP is assumed to be driven by the AFM transition, while in the second case AFM is only found upon Cd doping. The full symbols indicate \Tc{} and \TN{} determined from specific heat measurements, for the samples presented in this study. Equivalent negative pressures were obtained by comparison of these transitions with literature values.}
\end{figure}

Following the work of Pham \textit{et al.} (Ref.~\citen{Pham2006}) single crystals of composition \CeCoInCd{} can be viewed as negative pressure of \CeCoIn . The actual doping is $x< 0.03$. See Supplementary Note~S.1, Supplementary Fig~S.1 and Supplementary Table~S.1 for a detailed characterization. The SC and AFM transition temperatures for samples with various cadmium concentrations were used to determine the position of the samples into the phase diagram and their corresponding negative pressure (Fig.~1). 
Samples of a corresponding negative pressure of $p=-0.7$\,GPa and $p=-1.45$\,GPa were used for the \muSR{} experiment while for the  XAS experiment the corresponding pressure was $p=0$\,GPa (pure system) and $p=-1.45$\,GPa. The neutron diffraction scans were recorded for the sample with $p=-0.7$\,GPa. The samples $p=-0.7$\,GPa and $p=-1.45$\,GPa exhibit a N\'eel temperature of \TN $=2.06(2)$\,K and \TN $=3.04(2)$\,K ,respectively, while superconductivity is found below \Tc $=1.95(2)$\,K and \Tc $=0.52(2)$\,K, respectively. 

\subsection{Magnetic order coexisting with superconductivity.} 

\begin{figure}
\centering
\makebox[\textwidth][c]{\includegraphics{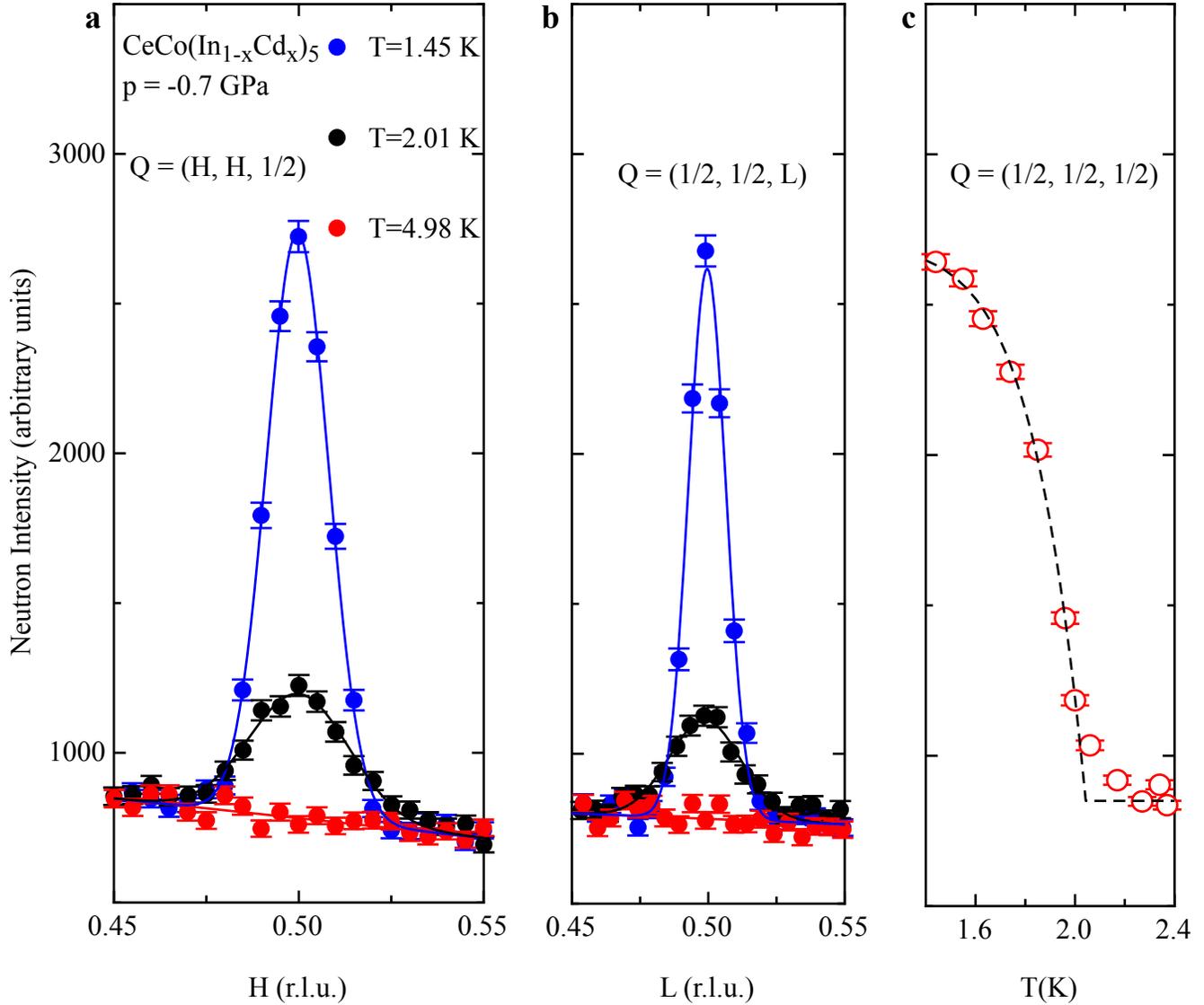}}
\caption{\label{Figure_Neutrons} \textbf{Scans in the reciprocal space of the neutron diffraction intensity of \CeCoInCd,} performed along the lines $\bf{Q}$=(H, H, 1/2) in (\textbf{a}) and $\bf{Q}$=(1/2, 1/2, L) in (\textbf{b}). Reciprocal lattice units (r.l.u.) are used as coordinates of the reciprocal space. The \CeCoInCd{} sample has a doping $x$ corresponding to a negative pressure of $p=-0.7$\,GPa. (\textbf{c}) Temperature dependence of the neutron diffraction intensity measured at $\bf{Q}$=(1/2, 1/2, 1/2).}
\end{figure}

The search for magnetic Bragg peaks was conducted using the neutron diffraction method. The neutron diffraction intensity was recorded in the direction corresponding to the lines ($H$, $H$, 0.5) for 0.4 $\leq H \leq$ 1 and (0.5, 0.5, $L$) for 0.4 $\leq L \leq$ 1 of the reciprocal space, below (1.45\,K and 2.01\,K) and above (4.98\,K) \TN . All the known propagation vectors in magnetically ordered compounds related to CeCoIn$_{5}$ are located on these lines. A single additional Bragg peak at $\bf{Q}$=(1/2, 1/2, 1/2) was identified below \TN , corresponding to a commensurate propagation vector $\bf{k}$=(1/2, 1/2, 1/2) (Fig.~2a,b). Since the position of the propagation vector $\bf{k}$ in the reciprocal space does not change with temperature, the temperature dependence of the neutron diffraction intensity at the center of the Bragg peak $\bf{Q}$ was recorded (Fig.~2c). Its variation is proportional to the square of the ordered magnetic moment.  The intensity of the Bragg peak at $\bf{Q}$=(1/2, 1/2, 1/2) vanishes above the N\'eel temperature of \TN $\approx$ 2.1\,K. This observation allows to relate the additional Bragg peak to the AFM order. The magnetic peaks are resolution limited at low temperatures ($T=1.45$\,K), the broadening of the peaks at $T$=2.01 K allows to estimate in-plane and out-of-plane correlation lengths to $\xi_{a}$(2.01~K) $\approx$ $\xi_{c}$(2.01~K) $\approx$ 140 $\AA$. Given the proximity to \TN , these are large values for $\xi_a$ and $\xi_c$.
The amplitude of the ordered magnetic moment can be estimated to be $m(p=-0.7$\,GPa$)\approx 0.4$\,$\mu_{B}$, which is consistent with the estimated moments of $m(p=-1.1$\,GPa$) \approx m(p=-1.7$\,GPa$) \approx 0.7$\,$\mu_{B}$ previously reported for higher cadmium concentrations\cite{Urbano2007,Nicklas2007}.

\begin{figure}
\begin{center}
\includegraphics[width=85mm]{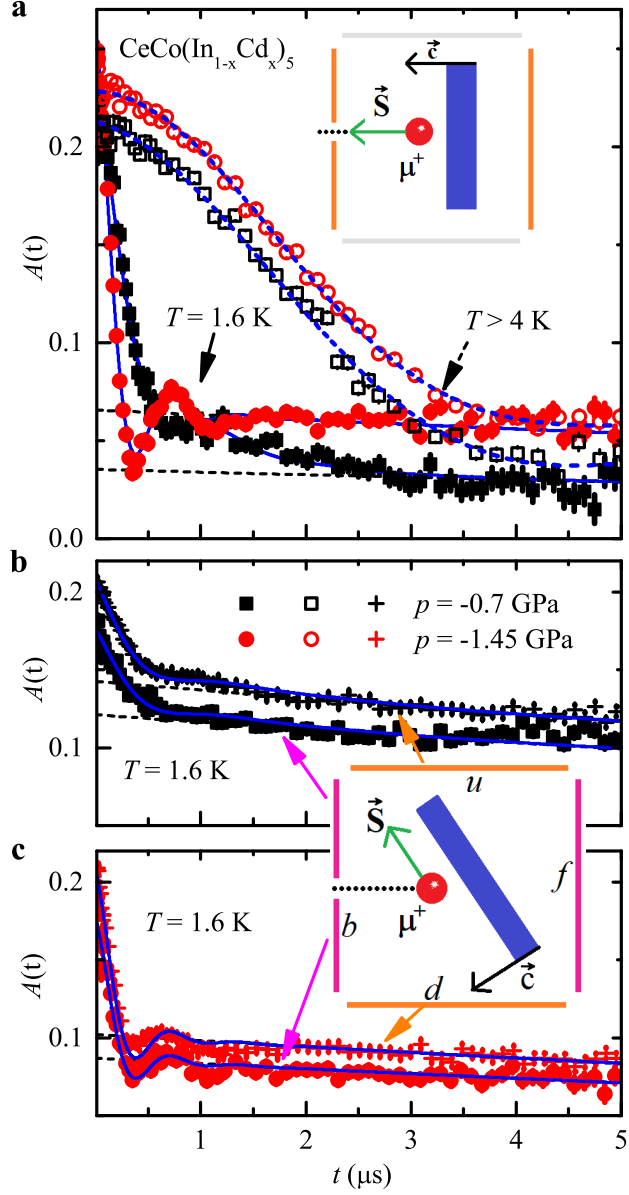}
\end{center}
\caption{\label{ZF} \textbf{$\bm \mu$SR zero field asymmetry spectra of \CeCoInCd.} Zero applied magnetic field \muSR{} asymmetry spectra have been recorded with the initial muon spin direction ($\vec{S}$) oriented parallel (\textbf{a}) and perpendicular (\textbf{b}\&\textbf{c}) to the crystallographic c-axis.
In panel (\textbf{a}) the signals above and below \TN{} are presented for \CeCoInCd{} with a doping inducing an equivalent chemical pressure of $p= -0.7$\,GPa (black) and $ p= -1.45$\,GPa (red). The oscillation in the asymmetry ($A(t)$) for $T=1.6$\,K indicates the presence of a long range magnetic order. In (\textbf{b}) and (\textbf{c}) the asymmetry corresponding to the two sets of detectors (crosses for up \textit{u} and down \textit{d} detectors relative to the initial muon momentum, full symbols for backward \textit{b} and forward \textit{f}) is presented for the two samples of \CeCoInCd{} composition and equivalent pressure $p=-0.7$\,GPa (\textbf{b}) and $p=-1.45$\,GPa (\textbf{c}).}
\end{figure}

Temperature dependent \muSR{} experiments were performed in zero applied external magnetic field ($\mu_0H=0$) and in an applied transverse field of $\mu_0H=50-54$\,mT. 
At 1.6\,K, in zero applied magnetic field, two different configurations of the initial muon spin relative to the sample crystallographic c-axis, sketched in Fig.~3 a and b, were investigated.
From the zero field \muSR{} asymmetry signal, the amplitude and the orientation of the magnetic field at the muon stopping site as well as the magnetic volume fraction were determined. The parameters magnetic fraction and moment orientation can be disentangle by the use of two different initial muon spin orientations, as indicated in eq.~\ref{eq:1}. At low temperatures, all the muons probe a local magnetic field at the muon stopping site. Therefore, the magnetic volume fraction reaches $\simeq100$\,\% at low temperatures in both samples, in agreement with the nuclear magnetic resonance (NMR) results\cite{Urbano2007}. For temperatures close to \TN{} in sample $p=-0.7$\,GPa, probably due to the intrinsic dopant distribution, the magnetic fraction is reduced and phase separation between paramagnetism and antiferromagnetism is observed (see Supplementary Note S.3 and Supplementary Fig.~S.2 for details).
The internal magnetic field at the muon stopping site is reduced by pressure, from $B_{m}(-1.45$\,GPa$)=11.2(1)$\,mT to $B_{m}(-0.7$\,GPa$)=6.5(2)$\,mT. It also rotates toward the $\vec{c}$ axis under chemical pressure. Similarly, both a reduction and rotation of the cerium magnetic moment were observed in the parent system CeRhIn$_5$ by nuclear quadrupole resonance under hydrostatic pressure\cite{Mito2001}.

\begin{figure}
\begin{center}
\includegraphics[width=85mm]{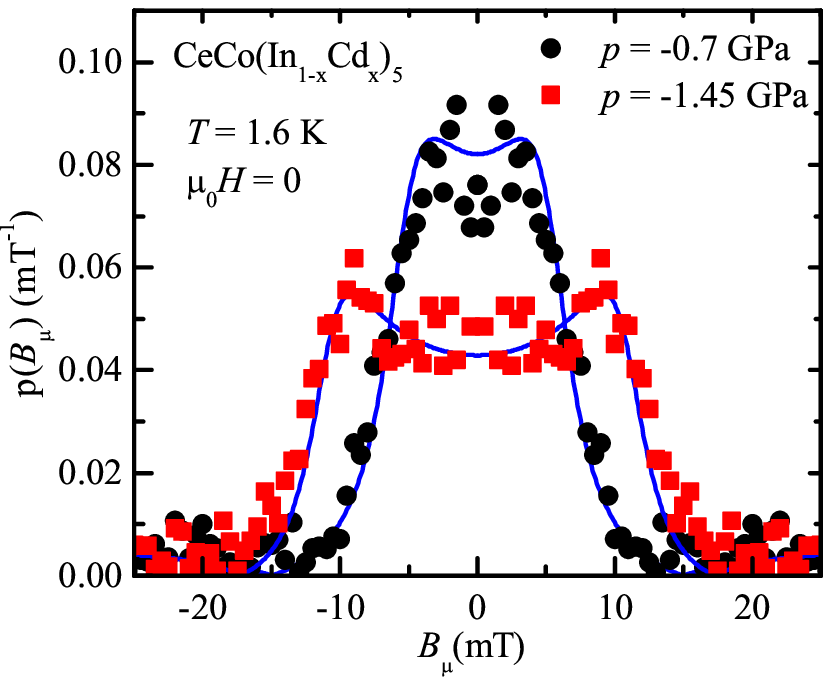}
\end{center}
\caption{\label{pB_LF} \textbf{Internal magnetic field distribution $\bm p(\bm B_\mu)$ at the muon stopping site in \CeCoInCd{}} for a doping inducing an equivalent chemical pressure of $p= -0.7$\,GPa (black) and $ p= -1.45$\,GPa (red). $p(B_\mu)$ is obtained by Fourier transformation of the magnetic contribution of the zero field \muSR{} asymmetry spectra ($A(t)$) at 1.6\,K of Fig.~3(\textbf{a}).}
\end{figure}

The Fourier transform of the \muSR{} asymmetry (Fig.~4 ) yields the magnetic field distribution at the muon stopping site $p(B_\mu)$. For commensurate magnetism, we would expect to observe a single internal distribution $p(B_\mu)$, or a discrete set of internal magnetic fields if there are several muon stopping sites. Here in contrary, independently of the chemical pressure, we observed a continuous distribution of internal magnetic field. The zero field \muSR{} spectra are best described by a Bessel type depolarization function (blue lines in Fig.~3), which corresponds for $p(B_\mu)$ to a broadened distribution of the form (blue lines in Fig.~4):
\begin{equation} \label{eq:pB_Bessel}
p(B_\mu)=1/(\pi\sqrt{B^2_{m}-B^2_\mu}).
\end{equation}

In \CeCoInCd{} a magnetic phase consisting of AFM droplets was proposed from NMR experiments\cite{Urbano2007,Seo2014,Grosche2014}. Droplets of fixed magnetic moment would form around cadmium impurity atoms, the magnetic fraction increasing with doping.
Such model appears highly unlikely regarding the internal magnetic field measured by \muSR{} (Fig.~4 ). In a droplet model we would expect a magnetic volume fraction of less than 100\%. Eventually, diluted microscopic droplets might induce a magnetic field in the surrounding non-magnetic regions, however in that case, we would expect the shape of the magnetic field distribution to change with doping. Furthermore, the value of $B_{m}$ should be independent of doping in a droplet model in contrary to the experimental result.

At first sight the muon and neutron studies appears to be incompatible. Indeed, the simplest magnetic structure leading to the observed neutron magnetic Bragg peak consists of moments localized on the cerium atom ordering with the propagation vector $\bf{k}$=(1/2, 1/2, 1/2) and forming a G-type antiferromagnetic order. In \CeCoIn{} two muon stopping sites were identified. The first one ($\simeq 70$\,\% of the muons) is located at position (0.5,0.5,0.5), i.e. in the center of a parallelepiped formed by eight cerium atoms\cite{Spehling2009}. The same position was found for CeRhIn$_5$\cite{Schenck2002}. At this position, no magnetic field will be generated by such simple magnetic structure, in contradiction with the \muSR{} results. Indeed, the first muon stopping site is a center of symmetry of the cerium crystallographic substructure of \CeCoInCd{} and regardless of their orientation magnetic moments located at the cerium site ordering with a propagation vector $\bf{k}$ cannot produce a magnetic field at such location. The second muon stopping site was identified to the position (0.5,0,0) or (0,0.5,0)\cite{Spehling2009} again in a localized G-type antiferromagnetic structure no magnetic field is produced at such position irrespectively of the moment direction at the cerium atom for symmetry reasons. The position of the muon stopping site is the main difference between \muSR{} and NMR that probe the two structurally different indium atoms. G-type antiferromagnetism can generate a magnetic field at both positions of the indium atoms, while it will not produce a magnetic field at the positions of the muon stopping sites.
The values $B_{m}=6.5$\,mT and $11.2$\,mT obtained in the \muSR{} measurements are not small magnetic field for this system, that could originate from defects in the magnetic structure. Indeed, the average magnetic field generated by a moment of $m=$0.4\,$\mu_B$ (0.7\,$\mu_B$ value of Refs.~\citen{Urbano2007,Nicklas2007} for samples of equivalent negative pressures $p=-1.1$\,GPa and $p=-1.7$\,GPa) on a sphere of radius $R=$5\,\AA{} (distance cerium--1$^\textrm{st}$ muon stopping site) is $B_{av} = 2\mu_0 m / (4\pi R^3) = 5.9$\,mT (10.3\,mT), a value comparable to $B_m$. Fluctuations of localized static magnetic moments could only generate a Gaussian distribution of field at each muons stopping site centered at $B=0$, with widths of a fraction of $B_{av}$.

However, both experiments appear to probe the same magnetic order. 
Indeed, both the amplitude and the pressure dependence of the internal field obtained in the \muSR{} experiment and the values of the average field generated by the magnetic moments of the intensity given by the neutron experiment are similar. 
Therefore, a different magnetic structure must exist that complies with the results of both experiments. 

\begin{figure}
\begin{center}
\includegraphics[width=85mm]{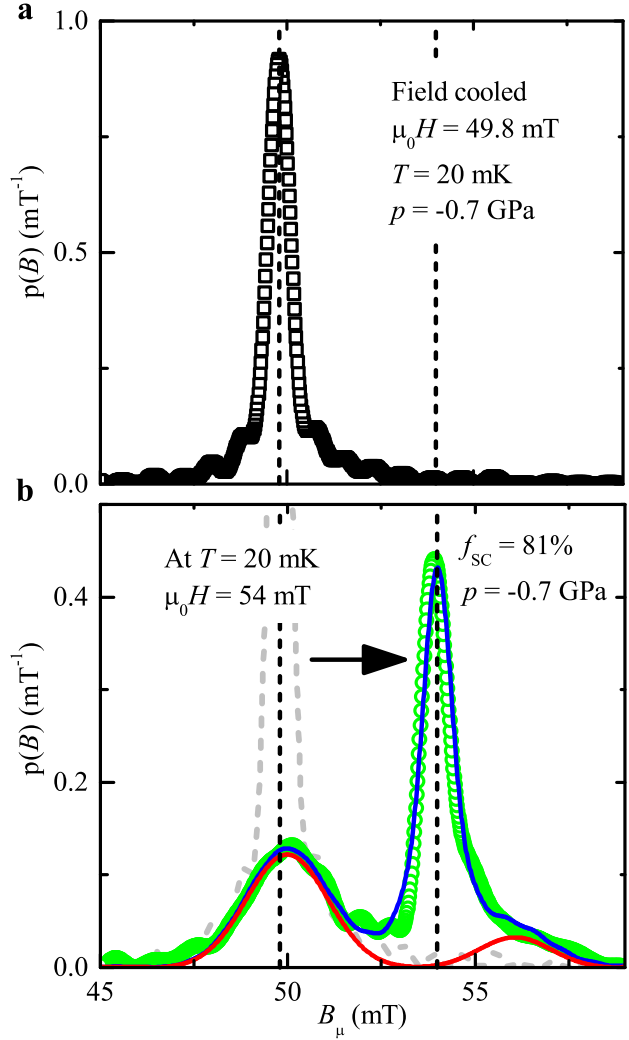}
\end{center}
\caption{\label{Bshift}\textbf{Vortex pinning in the SC phase of \CeCoInCd , $\bm p=\bm -0.7$\,GPa (field shift experiment).} The sample is cooled down to 20\,mK in an external magnetic field $\mu_0H\simeq 50$\,mT (field cooled), and the field distribution $p(B_\mu)$ is recorded [black squares in (\textbf{a}) and dashed gray line in (\textbf{b})] . The external magnetic field is then increased to $\mu_0H\simeq 54$\,mT, and the evolution of $p(B_\mu)$ is analyzed (\textbf{b}). Due to vortices pinning the magnetic field distribution in the SC fraction of the sample remains centered at a value $B\simeq 50$\,mT. In contrary muon stopping in the silver sample holder probe a narrow distribution of field centered at the value of the applied magnetic field. In (\textbf{b}) a third contribution is observed with muons probing a field distribution centered at  $\mu_0H\simeq 56$\,mT. It is attributed to muons stopping in a non-superconducting fraction of the sample.}
\end{figure}

The vortex state of \CeCoInCd{} is characterized by a strong vortex pinning\cite{Howald2013}, which was used to determine precisely the SC fraction of the sample $p=-0.7$\,GPa. The procedure consists in cooling the sample in an external field of $\mu_0H=49.8$\,mT down to 20\,mK. The field distribution at the muon stopping site is presented in Fig.~5a.
The external magnetic field is then increased to $\mu_0H=54$\,mT. The green circles in Fig.~5b represent the obtained field distribution. Three different contributions can be identified. In the SC region, the magnetic field distribution remains centered at $\mu_0H\simeq50$\,mT due to vortex pinning. The narrow distribution of fields centered at the applied field ($\mu_0H\simeq 49.8$\,mT or $54$\,mT) corresponds to muons stopping in the silver sample holder. In panel (\textbf{b}), a third contribution is obtained with a field distribution centered at $\mu_0H\simeq56$\,mT. A magnetic field larger than the applied one can be attributed to the contribution of superconducting stray field, originating either from muons stopping in the sample holder or in non-superconducting regions of the sample. In both cases, the muons would stop next to a superconducting region in the plane perpendicular to the applied magnetic field. An unpinned superconducting fraction is unlikely as the full superconducting fraction is pinned in pure \CeCoIn\cite{Howald2013} and cadmium dopants are more likely to increase the number of pinning center than to reduce the pinning potential. By comparison with other spectra, as the mean magnetic field of this contribution is large and the fraction of muons substantial we believe that this third contribution is due to a sample partially non-superconducting.
In Fig.~5b, the red line represents the fitted probability field distribution in the sample, while the blue line is the total fitted field distribution (sample + sample holder), assuming three independent Gaussian contributions. We found that $81(5)$\,\% of the muons stopping in the sample experience a magnetic field of $\simeq50$\,mT indicating pinning of the SC vortex lattice. This implies that at least  $81(5)$\,\% of the sample is SC. At this temperature the sample is also $\simeq$100\% magnetic, demonstrating \emph{phase coexistence between magnetism and superconductivity}. The remaining 19(5)\,\% of the sample volume is probably non-superconducting. It might indicate the formation of non-SC droplets around the Cd impurity atoms, in analogy to the conclusion of the NMR work\cite{Urbano2007} or be due to a small doping inhomogeneity (see Supplementary Note S.3). Non-SC droplets could also certainly explain the increase of the relaxation $T_{1}^{-1}$ in the superconducting state with cadmium doping ($x$) found in the NMR work\cite{Urbano2007}.

\subsection{Absence of cerium valence transition and lowering of the electronic Ce\,4$\bm f$ energy level.}

\begin{figure}
\begin{center}
\includegraphics[width=85mm]{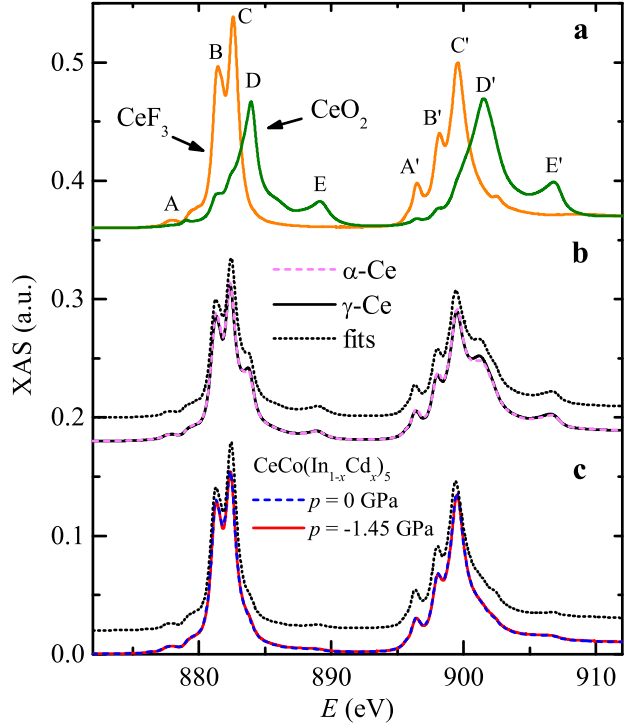}
\end{center}
\caption{\label{GraXAS} \textbf{Valence of different cerium based compounds.} M-edges XAS spectra were recorded for: (\textbf{a}) CeF$_3$ (orange), CeO$_2$ (green), (\textbf{b}) $\alpha$-Ce (magenta), $\gamma$-Ce (black ) and (\textbf{c}) \CeCoInCd{} of corresponding hydrostatic pressure $p=0$ and $p=-1.45$\, GPa respectively. The two systems in (\textbf{a}) have a valence of 3+ and 4+, respectively, allowing to determine the intermediate valence state of the metallic cerium systems and \CeCoInCd{} systems to be $\simeq 3.38$ and $\simeq 3.15$, respectively [dotted lines in (\textbf{b}) and (\textbf{c})]. Some curves are shifted vertically for clarity.}
\end{figure}

X-ray Absorption Spectra (XAS) of the M$_\textrm{4,5}$ edges (electron excitations from 3$d$ to 4$f$ angular momentum level) were recorded for several cerium based systems. In Fig.~6a, typical XAS spectra for the two valence configurations Ce$^\textrm{3+}$ and Ce$^\textrm{4+}$ are shown, using powders of the Mott insulator system CeF$_3$ and of the band insulator system CeO$_2$, respectively (see Supplementary Note S.4 and Supplementary Fig.~S.3 for an estimate of the exact valence). Spin orbit coupling of the 3d hole splits the spectrum in two edges: M$_\textrm{4}\approx 900$\,eV and M$_\textrm{5}\approx 882$\,eV.
The substructures are mainly given by the atomic multiplet structure which is clearly different between the valences Ce$^\textrm{3+}$ and Ce$^\textrm{4+}$.
Peaks A to C and A' to C' come from the Ce$^\textrm{3+}$ multiplet while peaks D, E, D' and E' come from the Ce$^\textrm{4+}$ multiplet.
Using these features, comparison can be done with pure metallic cerium in the $\gamma$-phase (room temperature) and the $\alpha$-phase ($\simeq 4$\,K) (Fig.~6b) as well as pure and cadmium doped \CeCoIn{} (at $\simeq 4$\,K Fig.~6c). The direct comparison is possible due to the absence of dichroism effects. No dichroism is present in the powders and in the amorphous cerium metal as all directions are averaged. The absence of dichroism in \CeCoIn{} was reported in \cite{Willers2010} and confirmed during the present experiment.

We obtained that both the $\alpha$ and $\gamma$ metallic phases of cerium are composed of $\nu_f\simeq 62$\,\%  of Ce$^\textrm{3+}$, or a valence of $\simeq$3.38. The variation of valence between the two phases is less than 0.03. We cannot exclude the formation of an oxide layer on the metallic cerium during the few minutes needed to transfer the sample into the chamber, but the value of $\nu_f$ is in good agreement with the 57\,\% of Ce$^\textrm{3+}$ found previously in the $\alpha$ phase on an in-situ evaporated cerium film\cite{Ottewell1973}. The absence of valence variation is in agreement with Compton scattering experiments\cite{Kornstaedt1982}. A positron annihilation experiment has revealed a valence in the range 3-3.5 for both $\alpha-$ and $\gamma-$ Ce phases\cite{Gustafson1969}. 
 
Similarly, a value of $\nu_f=85$\,\% Ce$^\textrm{3+}$ or a valence of $\simeq 3.15$ is obtained for the two \CeCoIn{} systems. The variation of valence with chemical pressure is less than 0.02. On the pure system a magnetic field up to $\mu_0H=6.5$\,T in the orientation $\vec{H}\parallel\vec{c}$ was applied without modifying the valence. See Supplementary Note S.5, Supplementary Table~S.3 and Supplementary Fig.~S.4 for details. For \CeCoIn{} both the valence value and the small influence of doping are in agreement with previous experiments\cite{Willers2010,Booth2011,Dudy2013}, even though the doping was different in those studies (Co$\rightarrow$Rh,Ir or Ce$\rightarrow$Yb).

As often discussed in the literature (see e.g.~\citen{Gustafson1969}) the absence of strong valence variations invalidate the promotional model for the transition between the $\alpha$ and $\gamma$ phases of metallic cerium. Similarly, no strong variation of valence occurs with doping in \CeCoInCd . The absence of valence modification across the field induced QCP ($\mu_0H_{QCP}=4.8$\,T, Ref.~\citen{Howald2011}) in \CeCoIn{} is particularly interesting as it allows to \emph{dismiss any direct influence of the valence (promotional like model or valence fluctuations model) on the Ce\,4$f$ ``delocalization'' leading to the suppression of the AFM phase.} % effect leading to the suppression of antiferromagnetism.} 
Indeed, \CeCoIn{} ($p=-1.45$\,GPa $H=0$) is AFM at low temperature while \CeCoIn{} ($p=0$\,GPa $\mu_0H=6.5$\,T) with the same valence is paramagnetic. The magnetic ground state of \CeCoIn{} ($p=0$\,GPa $H=0$) is unclear due to the occurrence of superconductivity.

\begin{figure}
\makebox[\textwidth][c]{\includegraphics[width=180mm]{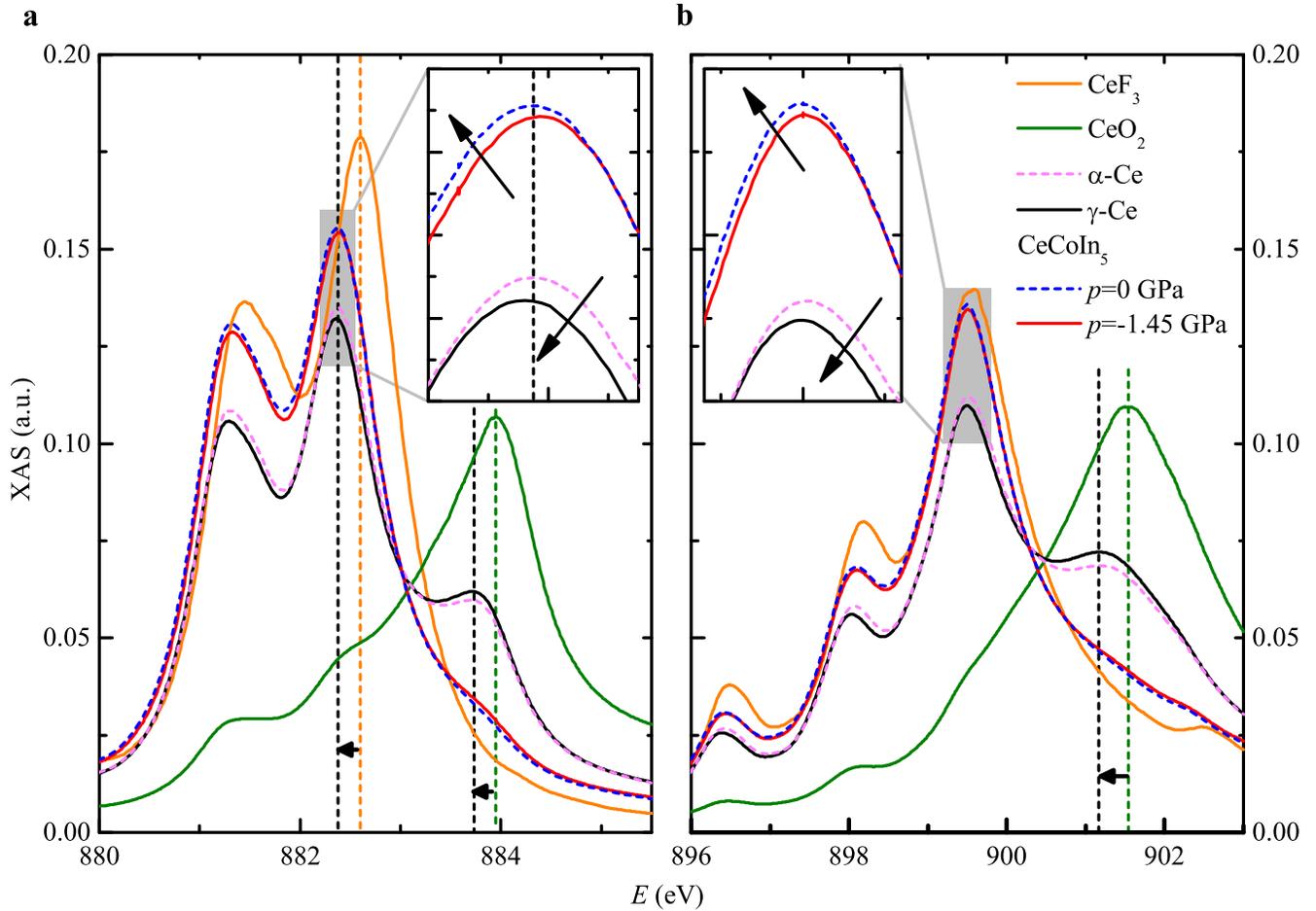}}
\caption{\label{GraXAS_M} \textbf{Energy lowering of the M-edge in metallic cerium and \CeCoInCd .} Details of the cerium XAS M$_\textrm{5}$ (\textbf{a}) and M$_\textrm{4}$ (\textbf{b}) edges (Fig.~6) are presented. Compared to the insulator references CeF$_3$ and CeO$_2$ the energy of the edges of the metallic samples is reduced by $\simeq 150$\,meV, as indicated by the vertical dashed lines. The cerium $\alpha$--$\gamma$ transition and the evolution of \CeCoIn{} with cadmium doping are characterized by a small variation in the cerium valence (peak height) and in the energy of the M$_\textrm{4,5}$ edges (peak position) in opposite directions. The pattern enlarged in inserts is observed consistently on each peak. The error bars on the data in insets are of the size of the lines thickness, see Supplementary Note S.6 and Supplementary Fig.~S.5.} 
\end{figure}

Figure~7 offers a closer view of the two absorption edges for the different systems. The four metallic systems display a systematic energy lowering of the absorption peaks compared to the insulator references, as indicated by the dashed vertical lines. 
Such an energy shift can either be due to an increase in the energy of the initial 3$d$ state or a decrease of the energy of the final 4$f$ state. 
Modification of coordination chemistry or in other words of the dispersion of the electron cloud surrounding each atom will modify the charge screening of the cerium ion. The variation in Coulomb interaction affects the energy of the surrounding electrons and particularly the ones on core orbitals such as the initial 3$d$ state. In insulator systems such flow of charge is directly related to the ion valence. A linear relation between the absorption edge energy shift and the valence was found at the vanadium K-edge\cite{Wong1984}, the rhenium L-edge\cite{Ravel2010} and the cerium L-edge\cite{Sarode1982}, using for each element different alloys to obtain different valences. In the two first systems, the energy shift of the pure metallic element follows the same linear relation, assuming an ionic screening or equivalent valence of 0, corresponding to the formation of an electronic band. 

\begin{figure}
\begin{center}
\includegraphics[width=85mm]{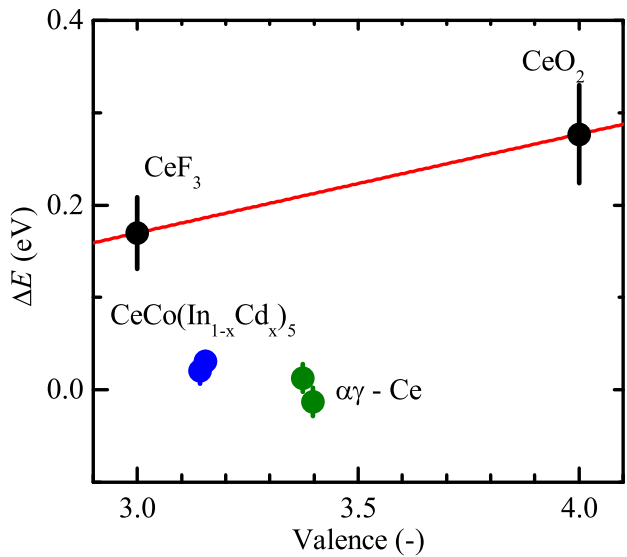}
\end{center}
\caption{\label{Eshift} \textbf{Energy shift in the XAS M$_\textrm{4,5}$-edge for different cerium based systems.} The average difference in energy of the M$_\textrm{4,5}$ absorption edges of each system compared to the average value of metallic cerium is plotted versus the valence for CeF$_3$, CeO$_2$ (black), \CeCoIn{}, \CeCoInCd{} $p=-1.45$\,GPa (blue), metallic cerium in the $\alpha$ and $\gamma$ phase (green). The red line represent the shift expected due to the variation in ionic screening with the modification of the valence. The two metallic cerium phases and \CeCoInCd{} are clearly located below this line, suggesting an at least partial band formation of 4$f$ angular momentum character.} 
\end{figure}

Taking the derivative of the XAS spectra, the energy position of the maximum of each absorption peak was determined. The shift of energy of each peak compared to the average value of the metallic cerium systems is then calculated. As this shift is found to be independent from the absorption peak energy (A-E, A'-E') the average value and standard deviation are plotted in Fig.~8 versus the valence. The energy shift expected due to the variation of ionic screening for different valences is extracted from the insulators systems (red line in Fig.~8). The two metallic cerium phases and \CeCoInCd , are clearly located below this line, and the shift is not directly related to the valence. In analogy to the case of vanadium and rhenium, this can be attributed to a stronger variation of the ionic screening due to the formation of a metallic band of partial Ce\,4$f$ angular momentum character.

Similarly, a weak hybridization of the Ce\,4$f$ final state, either due to the Kondo effect or to the formation of 4$f$ metal ligands (Mott--like delocalization) would result in a shift of the M$_\textrm{4,5}$-edge to lower energy. Such interpretation is corroborated by the observation of a comparable decrease $\simeq 0.2$\,eV of the 4$f$ state energy level in CeCoIn$_5$ compared to CeRhIn$_5$ in a x-ray photoemission experiment\cite{Treske2014}. CeRhIn$_5$ is usually considered as fully localized. 
We note that in pure cerium the Kondo energy must be substantially different between the $\gamma$ and $\alpha$ phases, as the effective quasiparticles mass strongly varies between $m^\star\simeq 6 m_0$ in the $\gamma$-phase and $m^\star\simeq 20 m_0$ in the $\alpha$-phase\cite{VanderEb2001}. If the energy shift was due to the Kondo effect we would expect a substantially different shift between the $\alpha$ and the $\gamma$ phase, in contrary to the experimental observations (Fig.~8). As the value of the shift is very similar in the \CeCoInCd{} systems it is likely that the same effect is responsible for the shift in all four systems.

The two mechanisms that might decrease the energy of the absorption edge imply the formation of a narrow electronic band of partial Ce\,4$f$ angular momentum character. We therefore conclude that the presence of Ce\,4$f$ metal ligands is the main reason leading to the energy shift in the M$_\textrm{4,5}$-edges XAS of $\alpha$ and $\gamma$-phase cerium and in \CeCoInCd . Similarly, scanning tunneling spectroscopy experiments\cite{Aynajian2012,Allan2013} have found indications for a narrow band of strong 4$f$ electronic character in \CeCoIn{} which is absent in CeRhIn$_5$.

A closer look at the different XAS spectra of pure cerium (insets Fig.~7) reveals that the $\alpha$-phase is slightly more localized than the $\gamma$-phase both in terms of valence ($\nu(\alpha)<\nu(\gamma)$) and hybridization ($dE(\alpha)<dE(\gamma)$). In contrary, the valence is higher in the cadmium doped than in the pure \CeCoIn{} systems, while the hybridization is increased. The behaviour enlarged in insets of Fig.~7 is observed consistently on each peak. A quantitative estimate of these variations is given in Table~S.3 using Eq.~(\ref{Eq:Valence}), assuming a fixed continuum contribution. Such opposite behaviours, indicate that the nature of the magnetic transition in \CeCoInCd{} under chemical pressure and between the cerium $\alpha-$ and $\gamma-$ phases is different.

\section*{Discussion}
Due to theoretical considerations it is often assumed that Ce\,4$f$ orbitals do not directly participate to metallic bonding. The reason is that the radius of maximum electronic density for Ce\,4$f$ orbitals is a factor $\approx 4$ smaller than the one of Ce\,5$d$6$s^2$ orbitals. In a compound, the average distance between the cerium and other nuclei is dominated by the Ce\,5$d$6$s^2$ character bands formation and therefore bands of partial Ce\,4$f$ orbital momentum character are assumed to have a higher energy than localized Ce\,4$f$ orbitals (see e.g. Ref.~\citen{Nikolaev2012}).

The large coherence length observed in the neutron diffraction experiment suggests homogeneous magnetic moments, which indicate that each site has the same mixed valence. This appears incompatible with a fully localized picture of Ce\,4$f$ electrons in which the valence of individual atoms can only take integer values.
The \muSR{} and neutron scattering experiments probe the magnetic order on different time and length scales. Indeed long range magnetic order is observed starting from a coherence of a few unit cell in the muon case, while the magnetic order revealed by the neutron Bragg peaks has a coherence length that exceed $100$\AA. One could therefore assume the existence of two separate magnetic orders. The first one, G-type commensurate localized antiferromagnetism would give no muon signal and the neutron Bragg peak, while the second incommensurate order, with short coherence length would be responsible for the muon signal. This scenario gives however no explanations why the two orders would have the same amplitude or why the short range magnetic order was not detected by neutron diffraction. A short range order would give a broader diffraction peak but should still be observed. A more convincing explanation is found by looking at the time scales of the two probes:
 $\tau\approx\mu$s for the muons, while $\tau\approx$\,ps for the neutrons. A magnetic order occurring in a band of partial Ce\,4$f$ character, such as the one revealed by the XAS measurements, is expected to have both localized and itinerant structures. On short time scales, the 4$f$ character dominates, with electrons found preferentially at the cerium position and domains with magnetic moment localized on the cerium site will be probed. Due to the electronic motion of the Ce\,4$f$ electrons, the size of these commensurate magnetism domains will shrink at larger time scales, and the incommensurate structure formed by the itinerant charge carriers will be probed. Such incommensurate magnetism is expected to generate a Bessel type depolarization function for the \muSR{} asymmetry signal, as observed experimentally\cite{Yaouanc2011}. The time scale of such crossover depends on the electronic motion and can vary between different systems. In \CeCoInCd{} we expect this time scale to be within the six orders of magnitude of difference in sensitivity between the neutrons and muons probes.

In this study we report three different indications for the presence of a band of partial Ce\,4$f$ character in \CeCoInCd . (i) The mixed valence of cerium %combined with the narrow magnetic Bragg peaks 
is incompatible with Ce$^{3+}$ 4$f$ electrons of full localized orbital character. (ii) The energy shift observed in XAS measurement, and (iii) the observed \muSR{} and neutron signals of the magnetic order are expected in case of a narrow electronic band of partial Ce\,4$f$ character. In \CeCoInCd , Ce\,4$f$ electrons are therefore simultaneously ``delocalized'' by ligand formation, as indicated by the presence of a band of partial Ce\,4$f$ character, by the Kondo effect, as indicated by the large quasiparticles effective masses, and through the mixed valence of the cerium atom. Under chemical or hydrostatic pressure one of these ``delocalization'' mechanisms is modified and leads to the disappearance of the AFM order.

For a valence fluctuations induced AFM phase transition we would expect a variation of valence of 0.1-0.7\cite{Watanabe2011,Kornstaedt1982} at the transition, incompatible with the experimental results both in \CeCoInCd{} and metallic cerium. 
As the effects on the XAS spectra of the cerium $\alpha$--$\gamma$ transition and upon cadmium doping in \CeCoInCd{} are different, we may ascribe them to the two other ``delocalization'' mechanisms: Kondo effect and electronic band broadening (Mott ``delocalization").
The strong variation of effective mass in metallic cerium between the $\alpha$ and $\gamma$ phases\cite{VanderEb2001} suggests a modification of the Kondo hybridization for this transition\cite{Murani2005}. By elimination we find that the suppression of magnetism in \CeCoInCd{} with applied chemical/hydrostatic pressure, is certainly due to the broadening of a band of partial Ce\,4$f$ character. 

In our previous \muSR{} work on \CeCoIn{} under pressure\cite{Howald2013} we had found a doubling of the SC carrier density, in a band of strong 4$f$ orbital momentum character, between 0\,GPa and 1\,GPa. This is compatible with the band broadening expected for an increased cerium--cerium hybridization under pressure. 
As in the ferromagnetic case, band antiferromagnetism is driven by a ``Stoner like'' criterion for antiferromagnetism\cite{Fritsche1998}. The magnetic order can be suppressed due to band broadening under pressure, while cadmium doping can both reduce the electronic bandwidth by ``negative'' chemical pressure and increase the hole density of state. Both effects are favourable for the realization of the ``Stoner like'' criterion for antiferromagnetism and would at some point lead to a phase transition. The equivalence between cadmium doping and hydrostatic pressure is in such a case naturally understood.

When the ``delocalization'' occurs via the Kondo effect, localized electrons are introduced into the main Fermi surfaces of low 4$f$ angular momentum character. In contrary, Ce\,4$f$ ligand formation or band broadening, constitute a band of substantial 4$f$ angular momentum character.
For fully itinerant systems such as \CeCoInCd , the interaction between the two ``delocalization'' mechanisms will produce an electron transfer between different Fermi pockets, reminiscent of the proposed Kondo breakdown scenario\cite{Si2001}, but also indicating that carrier concentration is an important parameter for the SC phase diagram in \CeCoIn{} possibly as important as in the case of high-\Tc{} cuprates.

In conclusion, we have investigated the antiferromagnetic state of \CeCoInCd{} combining three different experimental probes: neutron scattering, muon spin rotation and X-ray absorption spectroscopy.
Using neutron diffraction we confirmed the commensurate nature of the long range antiferromagnetic order at short time scales and the absence of secondary magnetic Bragg peak. We established the value of the magnetic moment for a different doping as the one reported in the literature ($\approx 0.4$\,$\mu_B$ for a sample with \Tc$\approx$\TN). We used muon spin rotation to probe the magnetic field at two centers of symmetry of the cerium sublattice. We measured a magnetic field that matches the expected average magnetic field generated by the magnetic moments of amplitude obtained in the neutron diffraction experiment.
We observed that the position of the absorption edge in X-ray absorption spectroscopy is reduced toward lower energies in the case of \CeCoInCd{} and metallic cerium compared to reference insulating materials. The presence of a narrow electronic band of partial 4$f$ character was identified as a possible reason for this energy shift. Depending of the timescale such a narrow electronic band can display either localized or itinerant properties corresponding to the neutron and muon signals.

The magnetic and superconducting volume fractions ($\simeq100$\,\% and $\simeq81$\,\% respectively) were extracted from the muon spin rotation experiment, directly assessing the microscopic coexistence of the two orders. The large magnetic fraction together with the evolution of the internal magnetic field strongly question the proposal of magnetic island for this system and suggests that the concept of non-superconducting island is more appropriate.

A detailed analysis of the mixed valence in \CeCoInCd{} demonstrated the absence of valence variation with doping and under magnetic field, notably across the quantum critical point. The evolution of fine structures in the X-ray absorption spectra indicate that the $\alpha-\gamma$ phase transition in metallic cerium has a different nature than the magnetic phase transition in \CeCoInCd . Assuming that the Kondo effect dominates the physic of metallic cerium we conclude that a broadening of the Ce\,4$f$ band is the main reason for the suppression of the magnetic order in \CeCoInCd .

These results provide a new approach on the superconducting coupling mechanism in this system by revealing the nature of an instability coexisting with superconductivity.

\begin{methods} % less than 3000 words (1520 now)

\subsection{Neutron scattering}
 measurements were carried out on the cold neutron three axis spectrometer IN12 located at ILL, Grenoble. The initial neutron beam is provided by a double focusing pyrolithic graphite (PG) monochromator. Higher order contamination is removed before the monochromator by a velocity selector. Diffraction measurements were carried out using a PG analyzer operated in a flat (non-focusing) geometry in order to reduce the background. The spectrometer was setup in W configuration with open-open-open collimations. The initial and final neutrons wave-vectors were $k_i=k_f$ = 2.6 $\AA^{-1}$, chosen in order to minimize the neutron absorption of Co, In and Cd. The single crystal sample, of equivalent negative pressure $p=-0.7$\,GPa and of dimensions: $\approx 5\times$7$\times$0.2 mm$^3$ is the largest of the pieces used for the \muSR{} experiments. It was mounted in a helium-4 cryostat with the [1, -1, 0] axis vertical, the scattering plane being thus defined by [1,1,0] and [0,0,1]. The notations used are as follow : The scattering vector $\bf{Q}$ is decomposed into $\bf{Q}$=${\bm \tau}$+$\bf{k}$, where ${\bm \tau}$ is a reciprocal lattice vector corresponding to a Brillouin zone center position and $\bf{k}$ is the propagation vector for a given magnetic structure. The Cartesian coordinates, $H$ and $L$, of the scattering vector $\bf{Q}$ are expressed in reciprocal lattice unit (r.l.u.) ($\bf{Q}$=($H$, $H$, $L$)). The Bragg peaks were fitted by a resolution limited Gaussian lineshape at 1.45 K and by a convolution of a Lorentzian with a resolution limited Gaussian at 2.01 K. The temperature variation of the magnetic intensity was fitted by the phenomenological function, $I \propto 1-(T/T_N)^{\alpha}$ with $T_{N}$=2.09(2) K and $\alpha$=6.4(3). To estimate the magnetic moment, the (3/2, 3/2, 1/2) Bragg peak was considered and its intensity was normalized to the structural Bragg peak reflection (1,1,0). The similar scattering geometry of these two peaks leads to comparable corrections factors (Lorentz factor, absorption). For the magnetic peak, the Ce$^{3+}$ magnetic form factor was considered.  The magnetic reflection (3/2,3/2,1/2) is relevant to address the two cases of magnetic moment direction given by group theory in a purely localized scenario ($c$-axis or in plane orientation). For these two possible spin orientations, the moment is found to be $\simeq 0.3 \mu_{B}$ ($c$-axis orientation) and $\simeq 0.5 \mu_{B}$ (in-plane orientation), whose average is given in the main text.

\subsection{Muon Spin Rotation (\muSR )}
experiments were performed at the Swiss Muon Source (S$\mu$S), Paul Scherrer Institut (PSI), Switzerland, using the GPS and the LTF spectrometers. In a \muSR{} experiment spin polarized positively charged muons are implanted into a sample. Due to their positive charge, the muons only stop at well defined locations in the unit cell, called the muon stopping sites. The initial muon spin orientation can be rotated by up to $\simeq 60^\circ$ before implantation in the sample of dimensions $\approx 8\times$12$\times$0.2 mm$^3$.
Within the present study, a mosaic of 8 ($p=-0.7$\,GPa) and 11 ($p=-1.45$\,GPa) single crystals were used.
In the presence of a local magnetic field at the muon stopping site $B_{\mu}$ the muon spin precesses at its Larmor frequency $\omega_{\mu} = \gamma_{\mu}B_{\mu}$, where $\gamma_{\mu}=135.54 \cdot 10^6$~HzT$^{-1}$ is the gyromagnetic ratio of the muon. The muon decays with a life time of $\tau_\mu=2.2$~$\mu$s into a positron and two neutrinos.
Due to parity violation the decay positron is preferentially emitted along the muon spin direction. Forward ($+$) and backward ($-$) positron detectors with respect to the initial muon spin polarization are used to monitor the \muSR{} asymmetry spectrum $A(t)$:
\begin{equation}
N_i(t)=N_{i0}\exp{(-t/\tau_\mu)}\left[ 1 \pm A(t)\right] + C_i
\end{equation}
where $N_i(t)$ is the positron count histogram on detector $i$. $N_{0i}$ and $C_i$ are constants.

In zero applied external magnetic field, the relaxation of $A(t)$ in the paramagnetic state is due to the nuclear moments and follows the so-called Kubo-Toyabe ($KT$) relaxation function\cite{Yaouanc2011}. In the presence of a static or quasi--static long range magnetic order at the muon stopping sites, $A(t)$ oscillates with time\cite{Yaouanc2011}.
For \CeCoInCd , the evolution of $A(t)$ is presented in Fig.~3a, with spectra taken above and below \TN .  While the frequency of the oscillations depends on the absolute value of the local magnetic field, the amplitude is related to the fraction and the relative orientation of $\vec{B_\mu}$. We may consider $A(t)$ as the sum of four different contributions:  The muons stopping in the silver sample holder give rise to a weak relaxation, visible at large time when contributions from the muons stopping in the sample have vanished (First term of eq.~\ref{eq:1}).  A potential non-magnetic fraction would lead to a $KT$ relaxation as observed above \TN (Second term of eq.~\ref{eq:1}). A magnetic fraction gives rise to two additional terms in $A(t)$ whose amplitude depends on the angle $\alpha$ between $\vec{B_\mu}$ and the initial muon spin orientation $\vec{S}$: an oscillating term for $\vec{S}\perp\vec{B_\mu}$ (Third term of eq.~\ref{eq:1}) and  an exponential damping for $\vec{S}\parallel\vec{B_\mu}$ (Fourth term of eq.~\ref{eq:1}).
\begin{equation}\label{eq:1}
\begin{alignedat}{2}
A(t) =& \\
&A_{Bg}\exp(-\gamma_\mu\Delta_{Bg}t) \\
+ &A_S (1-f_M) KT(\gamma_\mu\Delta_{KT},t) \\
+ &A_S f_M \\
& \qquad \left[ \langle\sin^2(\alpha)\rangle P_Z(t,B_{m}) \exp(-\gamma_\mu\Delta_P t) \right .  \\
& \qquad \left. + \langle\cos^2(\alpha)\rangle \exp(-\gamma_\mu\Delta_L t) \right] 
\end{alignedat}
\end{equation}
Here $A_i$ ($i=Bg$\& $S$) are initial asymmetries. $f_{S}=A_S/(A_S+A_{Bg})$ represents the fraction of muons stopping in the sample and $f_M$ is the magnetic fraction. $\Delta_{KT}$ and $\Delta_{Bg}$ are the field distributions due to nuclear moments in the sample and sample holder. The two other damping rates $\Delta_L$ and $\Delta_P$ are free parameters.  We have used a standard value of $\Delta_{Bg}=0.05$\,mT for the damping of the signal in the background, while $\Delta_{KT}\simeq 0.45$\,mT was found above \TN{} setting $f_M=0$ (empty symbols in Fig.~3a). Two different experimental configurations (inset Fig.~3) allow to implant muons with $\vec{S}$ either parallel ($\parallel$) or perpendicular ($\perp$) to the sample $\vec{c}$-axis.
Naming $\theta$ the angle between $\vec{B_\mu}$ and $\vec{c}$, we obtain $\alpha=\theta$ for the configuration $\parallel$ and $\langle\cos^2(\alpha)\rangle = (1 - \langle\cos^2(\theta)\rangle )/2$ and $\langle\sin^2(\alpha)\rangle = (1 + \langle\cos^2(\theta)\rangle )/2$ for the configuration $\perp$.  
Therefore, the asymmetries of the three different muon detectors configurations, schematically drawn in Fig.~3, can be fitted simultaneously using Eq.~(\ref{eq:1}).
The depolarization function $P_Z(t,B_{m})$ can be either a sum of cosines or a Bessel function depending on the magnetic structure, with $B_{m}$ the mean or respectively maximum field present at the muon stopping site.
The field distribution $p(B_\mu)$ obtained by Fourier transformation of the sample contribution to the asymmetry (Fig.~4 ) has clearly more the form of Eq.~(\ref{eq:pB_Bessel}) corresponding to the Bessel depolarization function than a sum of Dirac distributions centered at a finite magnetic field that would correspond to cosine depolarization functions. 
Detailed results of the fits performed in the time domain using the free software package MUSRFIT\cite{Suter2012} are presented in Supplementary Notes S.2 and S.3 and Supplementary Table~S.2.

\subsection{X-ray Absorption Spectroscopy (XAS)}
experiments were performed at the Swiss Light Source (SLS), Paul Scherrer Institut (PSI), Switzerland, using the X-Treme beamline, recording the total-electron-yield (TEY) current\cite{Piamonteze2012}.  The metallic samples of a typical size of 2x2x0.1\,mm$^3$ were glued with silver epoxy on a copper sample holder, cleaved or scratched in a Nitrogen atmosphere glove box before installation. The metallic cerium sample was annealed in vacuum at 750\,K before the experiment to insure a full $\gamma$ phase composition. The insulating powders were embedded in an indium foil to insure good electrical and thermal contact. In a XAS experiment, core level electrons are excited to the valence band by a monochromatic photon beam. When the core vacancies are refilled, Auger electrons are emitted, which scatter with other electrons in the material generating a TEY current between sample and ground. This current is normalized by the incoming beam intensity at each energy, monitored on a focusing mirror. The integral of the TEY spectra were normalized between 865\,meV and 920\,meV, after subtraction of the edge continuum (a Fermi step function centered at the M4-edge maximum was used). %, to get rid of the influence of surface quality and sample electrical conductance. 
Each spectrum is the average of 15 energy scans of 3 minutes each. The experiment was repeated at different sample positions without any significant modifications. None of the samples displays a linear dichroism as reported in Ref.~\citen{Willers2010}. The resolution is discussed in the Supplementary Note S.6.

In order to extract the valence of the metallic systems, the edges continuum was first subtracted on each XAS spectra. The spectra of CeF$_3$ (XAS$_\textrm{CeF3}(E)$) and CeO$_2$ (XAS$_\textrm{CeO2}(E)$) were then parametrized and used to fit the other spectra as:
\begin{equation}\label{Eq:Valence}
\textrm{XAS}_i(E) = \nu_f \cdot \textrm{XAS}_\textrm{CeF3}(E+dE) + (1-\nu_f) \cdot \textrm{XAS}_\textrm{CeO2}(E+dE)
\end{equation}
with $4-\nu_f$ the cerium valence of the analyzed system and $dE$ an energy shift. The fitted results are presented by dotted lines in Fig.~6 and the parameters reported in table~S.3.

\subsection{Time and length scales of the \muSR{} and neutron scattering experiments.}
The muon interaction time scale can be estimated using the precession frequency of the muon spin: $\gamma_\mu B_m \tau_\mu \approx 1$, $\rightarrow \tau_\mu \approx$ $\mu$s.
For a \muSR{} experiment a static or quasi--static long range magnetic order means that clusters ranging from a few unit cells to the whole sample, order statically on a time scale of a few $\mu$s. 
For the setup used in the neutron scattering experiment, the magnetic ordering is long range on the scale of at least 300\,\AA{} and static on the a scale of $\approx$\,ps, the neutron scattering interaction time scale.

\end{methods}

%\putbib[C:/Users/howald_l/Documents/Publi/biblio]
\end{bibunit}

\begin{addendum}
 \item We thank M.~Bendele for fruitful discussion as well as G.~Pascua for her participation to the XAS experiment. This work was mainly performed at the Swiss Muon Source (S$\mu$S) LTF and GPS spectrometers, at the Swiss Light Source (SLS) X-Treme spectrometer, Paul Scherrer Institut (PSI), Switzerland, and on the spectrometer IN12 of the Institut Laue--Langevin (ILL) in Grenoble, France. We acknowledge support by the Swiss National Science Foundation.
 \item[Competing Interests] The authors declare that they have no competing financial interests.
 \item[Correspondence] Correspondence and requests for materials should be addressed to L.H.~(email: ludovic.howald@psi.ch).
 \item[Author contributions]
GL grew the samples, GL and LH characterized them. SR performed the neutron experiment. CB, PDR and LH performed the \muSR{} experiments, PDR, AY and LH analyzed the results. ES, CP and LH performed the XAS experiment and analyzed the results. HK provides continuous experimental, scientific and financial support to the experiment. LH wrote the first version of the manuscript, all the authors contributed to the final version.
\end{addendum}

\beginsupplement

\newpage
{\spacing{1}\setlength{\parskip}{12pt}
\appendix{\Large\bfseries\sloppy
{\textsf{Supplementary information}}\\

\noindent \textsf{Evidence for Coexistence of Bulk Superconductivity and Itinerant Antiferromagnetism in the Heavy Fermion System CeCo(In$_{1-x}$Cd$_x$)$_5$}\\}
\setcounter{page}{1}

{\noindent\sloppy
Ludovic~Howald$^{1,2,\star}$, Evelyn~Stilp$^{1,3}$, Pierre~Dalmas~de~R\'eotier$^4$, Alain~Yaouanc$^4$, St\'ephane~Raymond$^4$, Cinthia~Piamonteze$^2$, G\'erard~Lapertot$^4$, Christopher~Baines$^3$ \& Hugo~Keller$^1$}
}

\begin{affiliations}
 \item Physik-Institut der Universit\"at Z\"urich, Winterthurerstrasse 190, CH-8057 Z\"urich, Switzerland
 \item Swiss Light Source, Paul Scherrer Institut, CH-5232 Villigen PSI, Switzerland
 \item Laboratory for Muon Spin Spectroscopy, Paul Scherrer Institut, CH-5232 Villigen PSI, Switzerland
 %\item Institut Nanosciences et Cryog\'enie, SPSMS, CEA and University Joseph Fourier, F-38054 Grenoble, France
\item Universit\'e Grenoble Alpes, INAC-SPSMS, F-38000 Grenoble, France and CEA, INAC-SPSMS, F-38000 Grenoble, France
\item[$\star$] ludovic.howald@psi.ch
\end{affiliations}

\begin{figure}[hb!]
\begin{center}
\includegraphics[width=85mm]{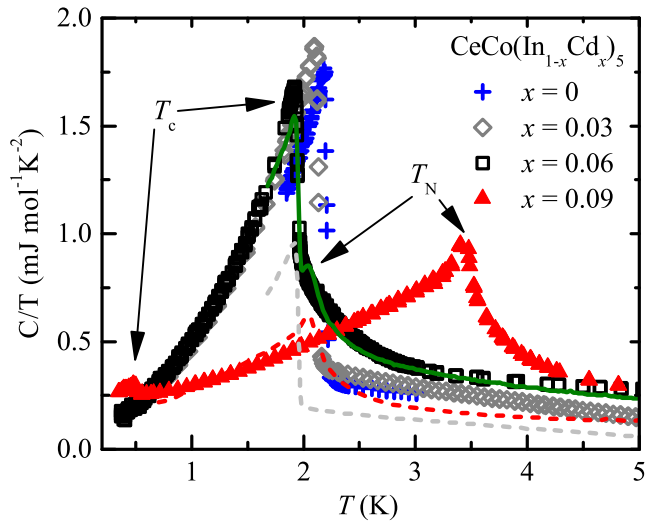}
\end{center}
\caption{\label{Cp} Temperature dependence of the specific heat in \CeCoInCd , $x$ stands for the nominal doping value. The step like transition indicates the onset of superconductivity (\Tc ), while the broader transition corresponds to the magnetic transition at the N\'eel temperature (\TN ). The green line is a sum of the parameterized curved ($x=0.03$ and $x=0.09$, dashed curves), scaled and shifted in temperature to fit the specific heat curve of the \CeCoInCd{} sample with $x=0.09$. It was used to determine \Tc{} and \TN{} for the sample of nominal doping $x=0.09$, as the features of the two transition temperatures overlap. 
}
\end{figure}

\begin{bibunit}
%\begin{refsection} % refsection environment
%
\Supsection{Sample characterization}\label{ap:Sample}\\
Single crystals of the tetragonal system \CeCoInCd{} were grown by the self--flux technique with different nominal cadmium concentrations ($x$)\cite{Canfield1992,MPC}. The actual cadmium concentration is much less\cite{Tokiwa2008} than the nominal concentration $x$ and depends on details of the sample growth. X-Ray Diffraction (XRD) indicates an actual concentration $\approx 10$ times smaller than the nominal one. Two types of transitions are observed in the temperature dependence of the specific heat (Fig.~\ref{Cp}). The sharp jump is characteristic of the SC transition (\Tc ), while the broader transition corresponds to the magnetic order (\TN ). The green line in Fig.~\ref{Cp} represents a fit discriminating the two transitions in sample $x=0.06$, using the shape of closely related samples ($x=0.03$ and $x=0.09$, dashed lines). Following the work of Ref.~\citen{Pham2006} the chemical pressure produced by the cadmium dopant in \CeCoInCd{} samples can be assimilated to a negative hydrostatic pressure ($p$) on the parent system \CeCoIn . Indeed, under hydrostatic pressure \CeCoInCd{} samples have a phase diagram very similar to \CeCoIn{} with a shift in the pressure scale of $\simeq0.14$\,GPa per percent of nominal doping $x$\cite{Pham2006,Gofryk2012}. To allow comparison with previous work the samples are named according to their corresponding hydrostatic negative pressures in this work. The corresponding hydrostatic pressure is determined by comparison of the AFM and SC transition temperatures to the ones reported in the literature. Transition temperatures and equivalent pressure for the samples presented here, are reported in table~\ref{SamplesTransitionT}.

\begin{table}[hb!]
\begin{center}
\begin{tabular}{c|c|c|c|c|c}
\hline
$x$ & \Tc{} (K) & \TN{} (K) & $p$ (GPa) & \muSR & XAS\\
\hline
0	& 2.21(2) & 0 & 0 & Ref.~\citen{Howald2013} & Yes\\
0.03 & 2.14(2) & $<0.3$ & -0.4 & -- & --\\
0.06 & 1.95(2) & 2.06(2) & -0.7 & Yes & --\\
0.09 & 0.515(15) & 3.44(2) & -1.45 & Yes & Yes\\
\hline
\end{tabular}
\end{center}
\caption{\label{SamplesTransitionT} SC and AFM transition temperatures for samples of various \CeCoInCd{} nominal compositions. Corresponding negative hydrostatic pressures and experiments performed are also indicated.}
\end{table}

\Supsection{Analysis of the \muSR{} spectra at 1.6\,K}\label{ap:muSR1p6}\\
\begin{table}[hb!]
\begin{center}
\begin{tabular}{c|c|c|c|c|c|c|c}
\hline
$p$ (GPa) & $B_m$ (mT) & $\theta$ (${}^\circ$) & $f_M$ (\%) & $\Delta_P$ (mT) & $\Delta_L$ (mT) & $f_{S\parallel}$ (\%) & $f_{S\perp}$ (\%)\\
\hline
-0.7 & 6.5(2) & 50(2) & 93(1) & 2.3(1) & 1.41(1) & 86(1) & 27(1)\\
-1.45 & 11.2(1) & 74(2) & 100(1) & 3.1(1) & 4.0(7) & 75(1) & 51(3)\\
\hline
\end{tabular}
\end{center}
\caption{\label{Parra} Parameters for the fit of the \muSR{} asymmetry of the  \CeCoInCd{} sample at 1.6\,K (Fig.~\ref{ZF}). $B_m$ is the amplitude of the maximum magnetic field at the muon stopping site, $\theta$ the angle between $\vec{B_\mu}$ and the sample $\vec{c}$-axis, $f_M$ the magnetic fraction and $f_{S(\perp/\parallel)}$ the fraction of muons stopping in the sample, for the two experimental configurations. $\Delta_P$ and $\Delta_L$ are the two damping rates defined in Eq.~(\ref{eq:1}).
}
\end{table}
The \muSR{} spectra in absence of a magnetic field were fitted using Eq.~(\ref{eq:1}) and a Bessel depolarization function. The results for the two different initial muon spin orientations ($\parallel$ and $\perp$) at 1.6\,K are presented in table~\ref{Parra}. The reduced stopping fraction in the sample $p=-0.7$\,GPa with configuration $\perp$ compared to the other sample is understood due to a sample holder with larger mount. The background contribution is represented by the dashed line in Fig.~\ref{ZF}. The parameters for the temperature dependence are presented in the next section.

\begin{figure}
\makebox[\textwidth][c]{\includegraphics[width=180mm]{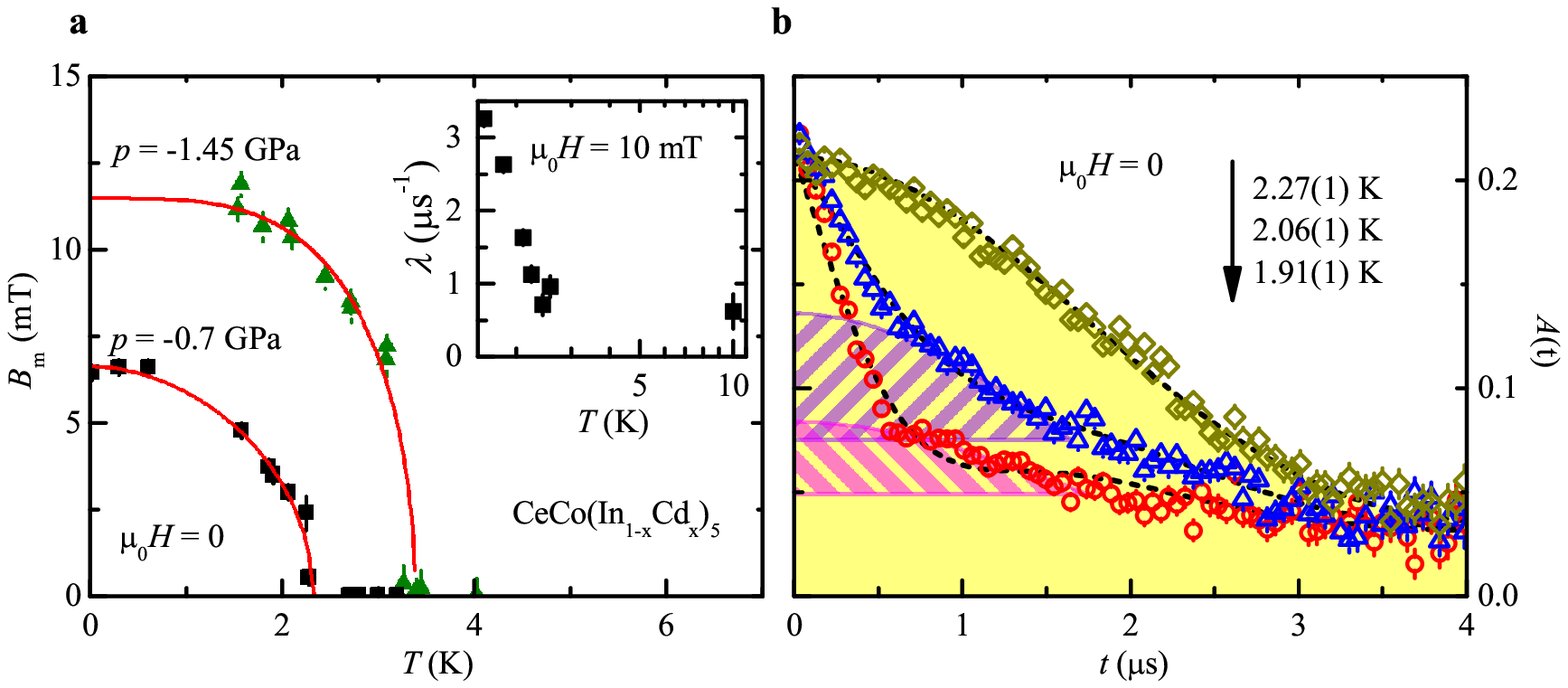}}
\caption{\label{ZF_LF} (\textbf{a}) Temperature dependence of the internal magnetic field for \CeCoInCd{} $p=-1.45$\,GPa and $p=-0.7$\,GPa. In the sample $p = -0.7$\,GPa a phase separation between an antiferromagnetic and non-magnetic (represented by the dashed areas) part is observed in the range 1.6\,K$<T<$2.3\,K (\textbf{b}). For the same sample the temperature dependence of the relaxation rate $\lambda$ in a longitudinal field is reported in the inset of panel (\textbf{a}).}
\end{figure}

\Supsection{Temperature dependence of the \muSR{} spectra}\label{ap:muSRTdep}\\
For the temperature dependence of the ZF \muSR{} spectra, only the configuration $\vec{S}\parallel\vec{c}$ was used. The \muSR{} asymmetry spectra were fitted globally with a simplified version of Eq.~(\ref{eq:1}) ($\alpha=90^\circ$). This simplification leads to a slight decrease in the fit quality. The perpendicular contribution is in such a model included in the background contribution: $\Delta_{Bg}$ was a free, temperature independent parameter. The AFM transition of the two \CeCoInCd{} samples ($p=-0.7$\,GPa and $p=-1.45$\,GPa) can be obtained following the temperature dependence of $B_{m}$ (Fig.~\ref{ZF_LF}a).

A temperature dependence of the form: $B(T)=B(0)\sqrt{1-(T/T_\textrm{N})^\alpha}$ was used. The values $\alpha=1.8(3)$ gives the best agreement for sample $p=-0.7$\,GPa. For the sample $p=-1.45$\,GPa we found $\alpha=3.7(7)$ and \TN $=3.38(2)$\,K in agreement with the specific heat transition (3.44(2)\,K). For the sample $p=-0.7$\,GPa the transition is more difficult to pinpoint as we observed a  phase coexistence in the temperature range 1.6\,K$<T<$2.3\,K. In this range, the \muSR{} asymmetry spectra (Fig.~\ref{ZF_LF}b) is the sum of a Kubo-Toyabe and an exponential decay contributions (Eq.~(\ref{eq:1}) with $0<f_M<1$). The non-magnetic contribution observed at 1.91\,K \& 2.06\,K is highlighted in Fig.~\ref{ZF_LF}b by the dashed area, that corresponds to a fraction of the normal state contribution (2.27\,K). In Fig.~\ref{ZF_LF}b, even a small non-magnetic fraction is clearly observable, as the Kubo-Toyabe function drastically differs from the Bessel depolarization function characteristic of the magnetic phase. In contrary, there is no signs of a Kubo-Toyabe contribution in Fig.~\ref{ZF}, indicating that the magnetic fraction is of $100$\,\% below 1.6\,K in sample $p=-0.7$\,GPa, and in the full temperature range investigated for sample $p=-1.45$\,GPa.

The fact that the temperature dependence of the internal field measure by \muSR{} differs from the temperature evolution of the amplitude of the magnetic moment measured by neutron diffraction, might be due to a larger doping distribution in the \muSR{} experiment than in the neutron diffraction experiment. At large length scale such as probed by neutron diffraction, the  average antiferromagnetic moment is reduced by the occurrence of superconductivity \cite{Nair2010}. This effect should differ on a length scale smaller than the superconducting coherence length such as probed by \muSR{} and is possibly another reason for the different temperature dependencies.

The transition is better observed in the temperature dependence of the damping parameter ($\lambda$) measured in longitudinal field ($\mu_0H=10$\,mT). The \muSR{} asymmetry spectra were fitted using a dynamic Kubo--Toyabe depolarization function. The temperature dependence of $\lambda$ is presented in inset of Fig.~\ref{ZF_LF}a. It gives \TN $\simeq 2.31$\,K, which is substantially higher than the specific heat transition (2.06(2)\,K), suggesting some distribution in the cadmium dopant concentration.  From the phase diagram (Fig.~\ref{Cp_phasediagram}), this doping distribution can be estimated to a maximum of $\delta p < 0.15\,GPa$ that corresponds to an effective doping variation of $\delta x<0.001$. Such a small doping distribution can occur between the different single crystals forming the sample mosaic or even within a single crystal.

\begin{figure}
\begin{center}
\includegraphics[width=85mm]{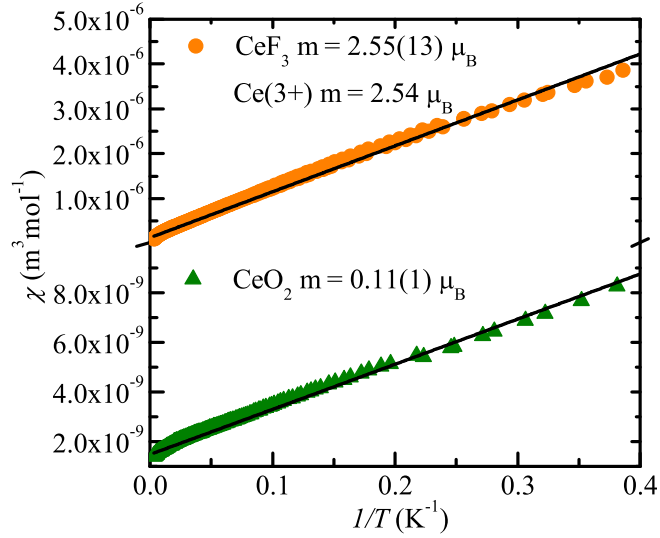}
\end{center}
\caption{\label{Xhi} Low temperatures Curie--Weiss susceptibility of the reference powder insulator systems CeF$_3$ and CeO$_2$. %The susceptibility of cerium in the $\alpha$-phase is shown for comparison.
}
\end{figure}

\Supsection{Valence of reference systems for XAS measurements}\label{ap:XASref}\\
The valences of the reference systems CeF$_3$ and CeO$_2$ were extracted via the temperature dependence of the susceptibility of the two systems.
Indeed, CeF$_3$ and CeO$_2$ do not have necessarily pure 3+ and 4+ valences  due to the possible formation of CeF$_4$ and Ce$_2$O$_3$ phases. The effective valences were measured via the moment of the cerium extracted from the Curie temperature dependence of the susceptibility: $\chi(T)=C/T$. The Curie constant has the form:
\begin{equation}
C=\frac{\mu_0\mu_B^2}{3k_B}N_A g_J J (J+1)
\end{equation}
where $\mu_0$ is the vacuum permeability, $\mu_B$ is the Bohr magneton, $k_B$ is the Boltzmann constant, $N_A$ is the Avogadro number, $g_J$ is the Land\'e $g$-factor, and $J$ is the angular momentum. The magnetic moment is obtained as: 
\begin{equation}\label{Eq:m}
m=g_J\sqrt{J(J+1)}\mu_B
\end{equation}
No moment is expected for cerium in the 4+ valence case $m_{4+}=0$ while for the 3+ contribution, Eq.~(\ref{Eq:m}) gives: $m_{3+}=2.54$\,$\mu_B$. The susceptibility of the two reference powders was measured in a SC quantum interference device (SQUID) (Fig.~\ref{Xhi}). We obtained that the reference systems CeF$_3$ and CeO$_2$ are at 100(1)\,\%, respectively 4(1)\,\% in the 3+ state. Including the deviation of CeO$_2$ from a pure 4+ state has no influences on the results discussed in the main article.

\Supsection{Parameters for the fits of the XAS spectra and evolution of the XAS spectra of \CeCoIn{} under magnetic field}\label{ap:XAS}\\
\begin{table}
\begin{center}
\begin{tabular}{l|cc|cc|c}
System & \multicolumn{2}{c|}{$\nu$} & \multicolumn{2}{c|}{$dE$ (meV)} & ``delocalization''\\
\hline
$\gamma$-Ce ($T=300$\,K) & 0.398(2) & \multirow{2}{*}{$\uparrow$} & 175(2) & \multirow{2}{*}{$\uparrow$} & \multirow{2}{*}{$\uparrow$}\\
$\alpha$-Ce ($T\simeq 4$\,K) & 0.374(1) & & 158(2) &  & \\
\hline
\CeCoIn{} $p=0$\,GPa & 0.142(1) & \multirow{2}{*}{$\downarrow$}& 153(1) & \multirow{2}{*}{$\uparrow$} & \multirow{2}{*}{$\uparrow$}\\
\CeCoIn{} $p=-1.45$\,GPa & 0.154(2) & & 145(2) & & \\
\end{tabular}
\end{center}
\caption{\label{TableValence} Evolution of the Ce\,4$f$ valence (3+$\nu$) and of the Ce\,4$f$ level hybridization ($dE$), obtained using Eq.~(\ref{Eq:Valence}). The values and errors are given assuming a fixed continuum contribution for each spectrum. The arrows indicate the observed physical ``delocalization'' (last column), as well as the one expected from the evolution of the parameters $\nu$ and $dE$.} 
\end{table}
\begin{figure}
\begin{center}
\includegraphics[width=85mm]{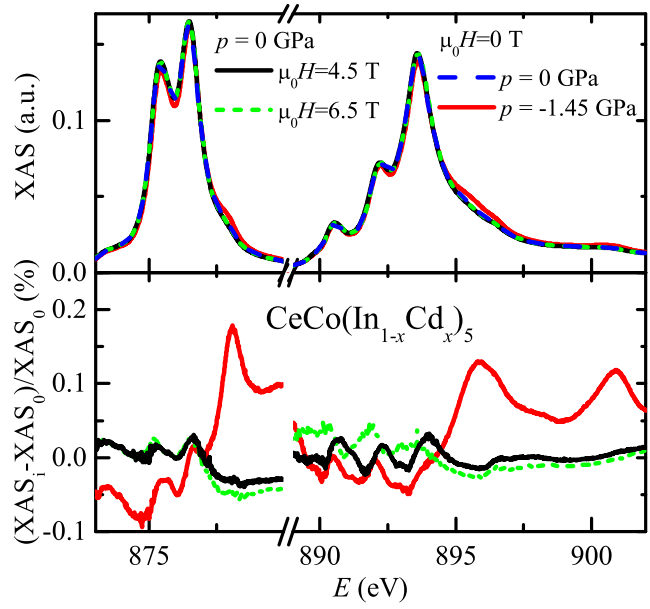}
\end{center}
\caption{\label{GraXAS_H} Evolution of XAS spectra in \CeCoInCd{} with doping and under magnetic field. In the bottom panel the normalized difference between the different spectra and a reference one $x=0$, $H=0$ at $T\simeq 4$\,K is plotted. The regions of low XAS absorption are not presented as the normalized difference between the spectra has less meaning.}
\end{figure}
In order to obtain the valence of $\alpha$ and $\gamma$ metallic cerium as well as of \CeCoInCd{} $p=0$\,GPa and $p=-1.45$\,GPa, their XAS spectra were fitted with Eq.~(\ref{Eq:Valence}). The results are displayed in table~\ref{TableValence}.

The absence of valence variation in \CeCoInCd{} with cadmium doping, does not directly imply that the magnetic transition is unrelated to a sizable variation of the valence, as the QCP possibly related to the AFM phase occurs under hydrostatic pressure\cite{Howald2011b} or magnetic field\cite{Howald2011}.
However, the magnetic field evolution up to 6.5\,T of the XAS spectra of pure \CeCoIn{} at low temperature ($\simeq 4$\,K) presented in Fig.~\ref{GraXAS_H} allows such a conclusion. In the lower panel the normalized difference to a reference spectra ($H=0$\,T $T\simeq 4$\,K) indicates the absence of strong valence variation across the field induced QCP ($H_{QCP}=4.8$\,T). The differences in XAS spectra between different magnetic fields are in fact smaller than for different doping. The absence of valence variation with cadmium doping in \CeCoInCd{} and under magnetic field in \CeCoIn{} indicates that the ``delocalization'' mechanism responsible for the AFM transition is not a valence transition.

\Supsection{Energy stability of XAS spectra}\label{ap:XASresolution}\\
\begin{figure}
\centering
\makebox[\textwidth][c]{\includegraphics[width=180mm]{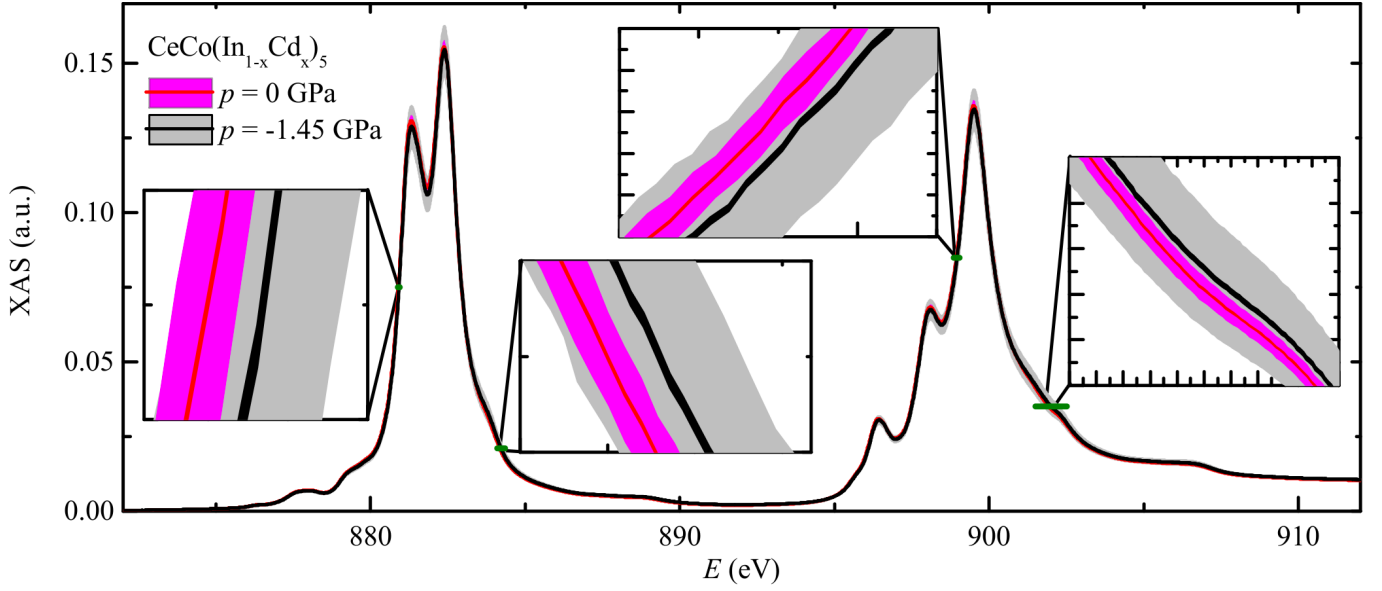}}
\caption{\label{ResolutionXAS} Resolution of the XAS experiment on \CeCoInCd . Insets are details of the XAS spectra at the position indicated by the black and green lines. Full grey and pink areas represent the standard deviation of the spectrum at each energy. Black and red areas represent the error in absolute energy position of the full XAS spectra.}
\end{figure}
The resolution of the instrument is given by the wavelength bandwidth of the photon beam, which is $\simeq 100$\,meV at the cerium edge.
The energy resolution only determines the size of structures in the spectrum which can be resolved. The energy shifts we observe do not have to do with energy resolution, but rather with the energy stability which measures how reproducible or reliable is the photon energy.
To quantify the energy stability, we calculated for each spectrum, using the 15 energy scans, the standard deviation at each energy (full grey and pink areas in Fig.~\ref{ResolutionXAS}).
As shown by the different zooms in Fig.~\ref{ResolutionXAS} the full curves are systematically shifted one relative to the other. 
The standard deviation in the position of the full XAS curve is therefore about the standard deviation of a single energy point divided by the square root of the total number of points with a significant energy shift (red and black area in Fig.~\ref{ResolutionXAS}). The difference between the two full XAS curves is statistically significant.

\vspace{3cm}

%\putbib[C:/Users/howald_l/Documents/Publi/biblio]

\end{bibunit}
%\printbibliography[heading=subbibliography]
%\end{refsection}

\end{document}